\documentclass[10pt, a4paper]{article}

\usepackage[a4paper, total={16cm, 25cm}]{geometry}
\usepackage{amsmath}
\usepackage{graphicx}
\usepackage{amsthm}
\usepackage{color}

\newtheorem{theorem}{Theorem}

\newtheorem{rational for conjecture}{Rational for Conjecture}

\begin{document}

\LARGE

\noindent \textbf{Estimation of Multiple Quantiles in Dynamically Varying Data Streams}

\begin{center}
  Hugo Lewi Hammer, Anis Yazidi and H{\aa}vard Rue
\end{center}

\normalsize

\begin{abstract}
In this paper we consider the problem of estimating quantiles when data are received sequentially (data stream). For real life data streams, the distribution of the data typically varies with time making estimation of quantiles challenging.

We present a method that simultaneously maintain estimates of multiple quantiles of the data stream distribution. The method is based on making incremental updates of the quantile estimates every time a new sample from the data stream is received. The method is memory and computationally efficient since it only stores one value for each quantile estimate and only performs one operation per quantile estimate when a new sample is received from the data stream. The estimates are realistic in the sense that the monotone property of quantiles is satisfied in every iteration. Experiments show that the method efficiently tracks multiple quantiles and outperforms state of the art methods.
\end{abstract}

\section{Introduction}

In this paper we consider the problem of estimating quantiles when data arrive sequentially (data stream). The problem has been considered for many applications like portfolio risk measurement in the stock market \cite{gilli2006application, abbasi2013bootstrap}, fraud detection \cite{zhang2008detecting}, signal processing and filtering \cite{stahl2000quantile}, climate change monitoring \cite{zhang2011indices}, SLA violation monitoring \cite{sommers2007accurate,sommers2010multiobjective}
and  back-bone network  monitoring \cite{choi2007quantile}.

Real life data streams typically have the following properties:
\begin{itemize}
\item[1.] The distribution of data from the data stream changes with time. All sorts of changes may happen like a shift of the distribution, change in the expectation value or the variance or other changes of the shape of the distribution.
\item[2.] Following a data stream over time, one may expect outliers, and some times extreme outliers.
\end{itemize}
In this paper we consider the problem of maintaining running estimates of multiple quantiles for data streams with the properties described above. A natural requirement of for the quantile estimates is that
\begin{itemize}
\item[3.] the monotone property of quantiles is satisfied, i.e. that the estimate of the, say, 70\% quantile always is above the estimate of the, say, 50\% quantile. 
\end{itemize}

\begin{figure}[h]
  \centering
  \includegraphics[width = \textwidth]{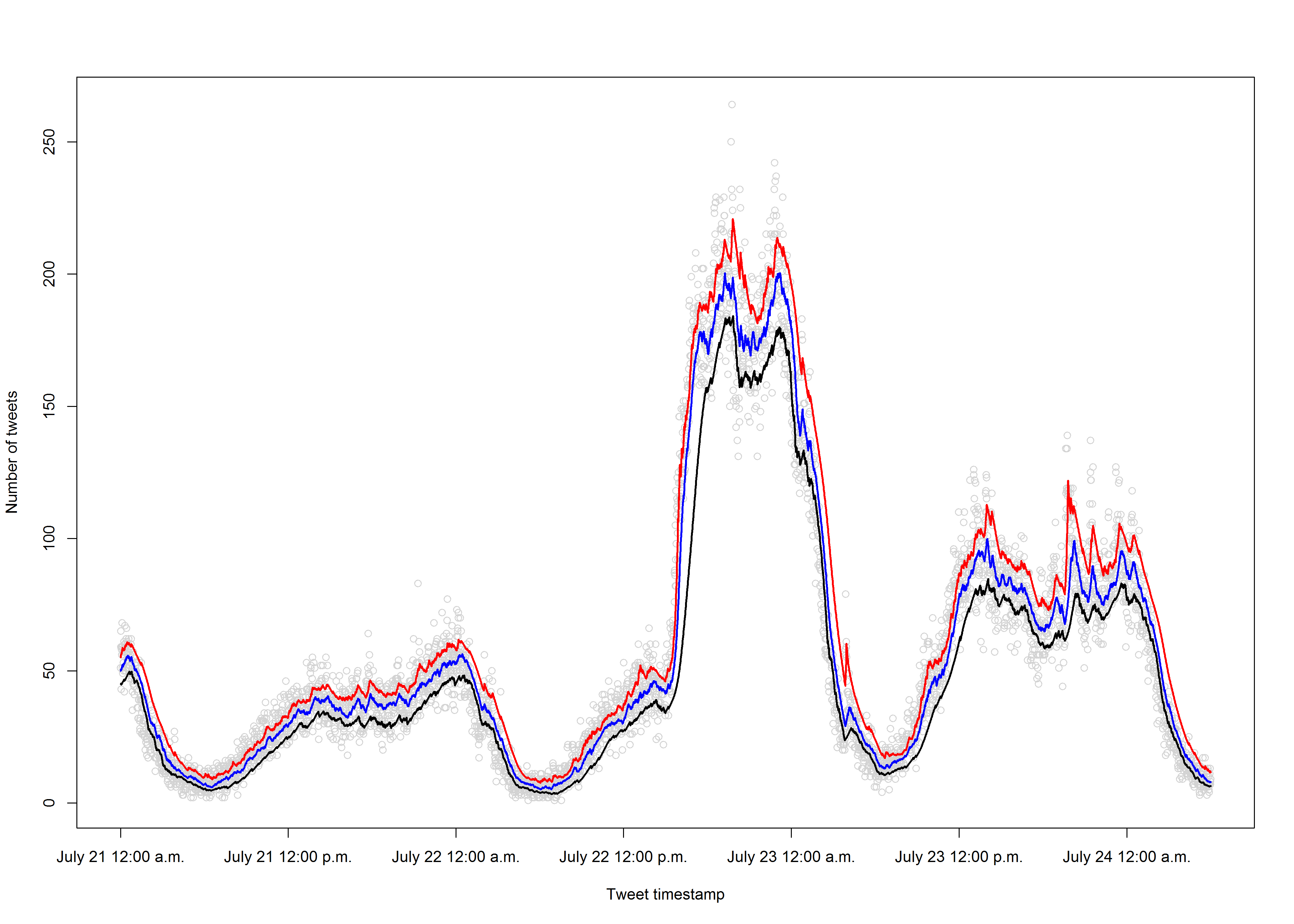}
  \caption{The gray circles show the number of tweets posted by Norwegian Twitter users every minute from July 21 2011 to July 24 2011. The black, blue and red curves show running estimates of the 20, 50 and 80\% quantiles of the distribution of the number of tweets posted.}
  \label{fig:6}
\end{figure}

Figure \ref{fig:6} shows an example to illustrate the problem.
The gray circles show the number of tweets posted by Norwegian Twitter users every minute in the time period before and after the Oslo bombing and Ut{\o}ya massacre in Norway July 22 2011. The terror attach started by a bomb going off in Oslo at July 22 3:25 p.m, and as expected, we see a rapid increase in the number of posted tweets after that time. The black, blue and red curves show the tracking of the 20, 50 and 80\% quantiles of the distribution of the number of tweets posted using the method that will be presented in this paper. We see that the method efficiently tracks the quantiles of the distribution, even when the distribution changes rapidly. We also see that the monotone property of quantiles is satisfied in every iteration.

Efficiently estimating quantiles for data streams with the properties described above (and as shown in Figure \ref{fig:6}), is a challenging task. The most natural is to maintain a sorted list of the data and estimate the quantiles from the sorted list. Such a quantile estimator has a clear disadvantage as computation time and memory requirement increases with time. In addition, the quantile estimates will not adapt to the dynamic changes of the data stream. An other alternative could be to fit a time series model to the received data and compute quantiles of the forecast distribution, but this is challenging since the dynamic changes of the distribution of the data stream is unknown.

Several quantile estimators have been proposed to estimate quantiles in dyanmically changing data streams. Most of the methods are based on keeping a representative sample, e.g. a sliding window, or summary information, e.g. a histogram, of the data. Quantiles are computed from these samples or summaries. See \cite{gilbert2002summarize,Arasu2004,gilbert2002fast,datar2002maintaining,lin2004continuously} for representative examples. Unfortunately, such summary information tend to be memory expensive and for data streams where the distribution is changing with time, this may lead to a large biases in the quantile estimates since most of the summary information may be out of date \cite{cao2009incremental}.

Another ally of methods are so called  \emph{incremental} update methods. The methods are based on performing small updates of the quantile estimate every time a new sample is received from the data stream. The methods only need to store one value for each quantile estimate and therefore are very memory efficient compared to the sample/summary methods described in the previous section. The literature on such quantile estimation methods is sparse. One of the first and prominent examples is the algorithm of Tierney (1983) \cite{Tierney1983} which is based on the stochastic approximation theory. The method is developed for a static data stream and will not work for dynamically changing data streams. A few modifications of the Tierney method have been suggested that are able to estimate quantiles for dynamically varying data streams, see e.g. \cite{Chen2000} and \cite{cao2010tracking}. For more recent methods we can mention the Frugal methods  \cite{ma2013frugal} which runs a discrete Markov chain and estimates quantiles of discrete probability distribution. Yazidi and Hammer (2016) \cite{yazidi16} proposed a version of the Frugal method that works on continuous sample spaces in addition to an improved version, based on deterministic updates. A disadvantage with the incremental methods referred to above is that they are constructed to estimate only \textit{a single} quantile of the data stream. Of course, one could run such methods for several quantile probabilities, but for such methods, the quantile estimates usually will be unrealistic since the monotone property of quantiles (item 3 above) usually will be violated. The problem with monotone property violation will be reduced if the incremental update size is reduced, but this is not a viable alternative for dynamically changing data streams. By using too small update steps, the incremental methods will not be able to track the dynamic changes of the data stream. In other words, a good quantile estimation algorithm must on one hand be able to efficiently track the quantiles of the data stream and at the same time satisfy the monotone property of quantiles in every iteration. The only method we have found in the literature that attempt to estimate multiple quantiles of dynamically changing data streams is the method of Cao et al \cite{cao2009incremental}. The method is based on first running an incremental update of each quantile estimate and secondly compute a monotonically increasing approximation of the cumulative distribution of the data stream distribution. Finally the quantile estimates are computed from the approximate cumulative distribution. A disadvantage of the method is that we do not have any guarantee that the resulting quantile estimates converges to the true quantiles. In addition, as part of the method, the data from the data stream will be used directly making it sensitive to outliers (recall property two of real life data streams).

In this paper we suggest a novel incremental method to estimate multiple quantiles. The method adapts efficiently to dynamically changing data streams and do not use the values of the data streams directly, but only if the values are above or below the current estimates. This makes the method robust to outliers. The method is constructed such that quantile estimates satisfy the monotone property of quantiles in every iteration. Therefore, the method suggested in this paper satisfies the requirements of real life data streams as described above. A theoretical proof will be given showing that the quantile estimates converges to the true quantiles.

\section{Estimation of multiple quantiles}
\label{sec:multq}

Let $X_n$ denote a stochastic variable for the possible outcomes
from a data stream at time $n$ and let $x_n$ denote a random sample (realization) of $X_n$.
 We assume that $X_n$ is distributed according to some distribution $f_n(x)$ that varies dynamically with time $n$.
 Further let $Q_{n}(q)$ denote the quantile associated with probability $q$, i.e $P(X_n \leq Q_n(q)) = F_{X_n}(Q_n(q)) = q$.

In this paper, we focus on simultaneously estimating the quantiles for $K$ different probabilities $q_1, q_2, \ldots, q_K$ at each time step (tracking) for a data stream where the distribution of the samples varies dynamically with time. We assume an increasing order of the target quantiles to be tracked, i.e. $q_1 < q_2 < \cdots < q_K$.
The straight forward approach to estimate the quantiles would be to run $K$ online quantiles estimators in parallel and in isolation, one for each probability. Using the DUMIQE approach from \cite{yazidi16}, the update equation reduces in this case to
\begin{align}
\label{eq:1}
  \begin{split}
   \widehat{Q_{n+1}}(q_k) &\leftarrow (1 + \lambda q_k) \widehat{Q_{n}}(q_k)  \hspace{14mm}\text{ if } \widehat{Q_{n}}(q_k) < x_n \\
   \widehat{Q_{n+1}}(q_k) &\leftarrow (1 - \lambda (1-q_k)) \widehat{Q_{n}}(q_k)  \,\hspace{5mm}\text{ if } \widehat{Q_{n}}(q_k) \geq x_n
  \end{split}
\end{align}
for $k = 1,2, \ldots, K$. We assume that $\widehat{Q_{n}}(q_k) > 0, k = 1,2, \ldots$. 
Unfortunately, using independent updates of each quantile results in unrealistic estimates as the monotone property of quantiles, as given by the constraint $\widehat{Q_{n}}(q_1) \leq \widehat{Q_{n}}(q_2) \leq \ldots \leq \widehat{Q_{n}}(q_K)$,
is most likely violated in some iterations. For the DUMIQE this can be explained as follows (see \cite{cao2009incremental} for an example of another method). Assume at time $n$ that the monotone property is satisfied and that the sample $x_n$ admits a value between $\widehat{Q_{n}}(q_k)$ and $\widehat{Q_{n}}(q_{k+1})$, i.e:
\begin{align}
\label{eq:2}
\widehat{Q_{n}}(q_1) \leq \cdots \leq \widehat{Q_{n}}(q_k) < x_n < \widehat{Q_{n}}(q_{k+1}) \leq \cdots \leq \widehat{Q_{n}}(q_K)
\end{align}
Then according to \eqref{eq:1} the estimates are updates as follows
\begin{align}
\label{eq:3}
  \begin{split}
  \widehat{Q_{n+1}}(q_j) &\leftarrow (1 + \lambda q_j) \widehat{Q_{n}}(q_j) \hspace{19mm}\text{ for } j = 1,2, \ldots, k\\
  \widehat{Q_{n+1}}(q_{j}) &\leftarrow (1 - \lambda (1-q_{j})) \widehat{Q_{n}}(q_{j}) \,\hspace{10mm}\text{ for } j = k+1, \ldots, K
  \end{split}
\end{align}
which means that the estimates are increased for the quantiles with an estimate below $x_n$ and decreased for the estimates above $x_n$. Consequently, the monotone property may be violated. In the next section, we will present a novel update scheme that satisfies the monotone property of quantiles while converging both theoretically and experimentally to the true quantiles.

\subsection{Multiple quantile DUMIQE}
\label{sec:md}

In order to maintain the monotone property, our idea is to adjust the value of $\lambda$ in \eqref{eq:1} to ensure that we satisfy the monotone property, i.e. $\widehat{Q_{n}}(q_k)$ should lay between $\widehat{Q_{n}}(q_{k-1})$ and $\widehat{Q_{n}}(q_{k+1})$ for every $k$ at every time step $n$. Let us assume at time $n$ that the monotone property is satisfied and that the sample $x_n$ gets a value between $\widehat{Q_{n}}(q_k)$ and $\widehat{Q_{n}}(q_{k+1})$ as given by \eqref{eq:2}.
A sufficient criterion to satisfy the monotone property between $\widehat{Q_{n+1}}(q_k)$ and $\widehat{Q_{n+1}}(q_{k+1})$ is to make sure to use a sufficiently small $\lambda$ such that $\widehat{Q_{n+1}}(q_k) \leq \widehat{Q_{n+1}}(q_{k+1})$. We find such a $\lambda$ (denoted $\widetilde{\lambda}$ below) by ensuring that the distance between $\widehat{Q_{n+1}}(q_k)$ and $\widehat{Q_{n+1}}(q_{k+1})$ is equal to some \emph{portion}, $\alpha$, of the distance from the previous iteration, i.e.
\begin{align}
  \label{eq:4}
  \begin{split}
  \widehat{Q_{n+1}}(q_{k+1}) - \widehat{Q_{n+1}}(q_k) &= \alpha \left(\widehat{Q_{n}}(q_{k+1}) - \widehat{Q_{n}}(q_k) \right) \\
  (1 - \widetilde{\lambda} (1-q_{k+1})) \widehat{Q_{n}}(q_{k+1}) - (1 + \widetilde{\lambda} q_k) \widehat{Q_{n}}(q_k) &= \alpha \left(\widehat{Q_{n}}(q_{k+1}) - \widehat{Q_{n}}(q_k) \right)
  \end{split}
\end{align}
By solving \eqref{eq:4} with respect to $\widetilde{\lambda}$ we obtain:
\begin{align}
\label{eq:5}
  \widetilde{\lambda} = (1-\alpha) \frac{\widehat{Q_{n}}(q_{k+1}) - \widehat{Q_{n}}(q_{k})}{(1-q_{k+1})\widehat{Q_{n}}(q_{k+1}) + q_k\widehat{Q_{n}}(q_{k})}
\end{align}
We must also ensure that $\widehat{Q_{n}}(q_{k})$ stays above the estimate below, $\widehat{Q_{n}}(q_{k-1})$. Thus a sufficient criterion to guarantee that $\widehat{Q_{n}}(q_k)$ stays between $\widehat{Q_{n}}(q_{k-1})$ and $\widehat{Q_{n}}(q_{k+1})$ is to use the minimum of $\widetilde{\lambda}$ computed from $\widehat{Q_{n}}(q_{k})$ and $\widehat{Q_{n}}(q_{k+1})$ and computed from $\widehat{Q_{n}}(q_{k})$ and $\widehat{Q_{n}}(q_{k-1})$. This gives the following:
\begin{align}
\label{eq:6}
  \begin{split}
  \widetilde{\lambda} &= \min \left\{ (1-\alpha) \frac{\widehat{Q_{n}}(q_{k}) - \widehat{Q_{n}}(q_{k-1})}{(1-q_{k})\widehat{Q_{n}}(q_{k}) + q_{k-1} \widehat{Q_{n}}(q_{k-1})}, (1-\alpha) \frac{\widehat{Q_{n}}(q_{k+1}) - \widehat{Q_{n}}(q_{k})}{(1-q_{k+1})\widehat{Q_{n}}(q_{k+1}) + q_k\widehat{Q_{n}}(q_{k})} \right\} = \\
&= (1-\alpha) \min \left\{ \frac{\widehat{Q_{n}}(q_{k}) - \widehat{Q_{n}}(q_{k-1})}{(1-q_{k})\widehat{Q_{n}}(q_{k}) + q_{k-1} \widehat{Q_{n}}(q_{k-1})}, \frac{\widehat{Q_{n}}(q_{k+1}) - \widehat{Q_{n}}(q_{k})}{(1-q_{k+1})\widehat{Q_{n}}(q_{k+1}) + q_k\widehat{Q_{n}}(q_{k})} \right\}\\
&= (1-\alpha) H\hspace{-1mm}\left(\widehat{Q_{n}}(q_{k}); \widehat{Q_{n}}(q_{k-1}), \widehat{Q_{n}}(q_{k+1}) \right)
  \end{split}
\end{align}
By using $\lambda = \widetilde{\lambda}$ from \eqref{eq:6} in \eqref{eq:1} when updating the estimates $\widehat{Q_{n}}(q_1), \widehat{Q_{n}}(q_2), \ldots, \widehat{Q_{n}}(q_K)$, we can guarantee that the monotone property will be satisfied for all the quantile estimates. Of course, the lowest quantile estimate $\widehat{Q_{n}}(q_1)$ only needs to satisfy the monotone property against $\widehat{Q_{n}}(q_2)$ and therefore $H$ becomes:
\begin{align*}
  H\hspace{-1mm}\left(\widehat{Q_{n}}(q_{1}), \widehat{Q_{n}}(q_{2}) \right) = \frac{\widehat{Q_{n}}(q_{2}) - \widehat{Q_{n}}(q_{1})}{(1-q_{2})\widehat{Q_{n}}(q_{2}) + q_{1} \widehat{Q_{n}}(q_{1})}
\end{align*}
and similarly for the highest quantile estimate:
\begin{align*}
  H\hspace{-1mm}\left(\widehat{Q_{n}}(q_{K-1}), \widehat{Q_{n}}(q_{K}) \right) = \frac{\widehat{Q_{n}}(q_{K}) - \widehat{Q_{n}}(q_{K-1})}{(1-q_{K})\widehat{Q_{n}}(q_{K}) + q_{K-1} \widehat{Q_{n}}(q_{K-1})}
\end{align*}
Substituting $\widetilde{\lambda}$ in \eqref{eq:6} for $\lambda$ in \eqref{eq:1} and defining $\beta = 1 - \alpha$, we obtain the following update rules:
\begin{align}
\label{eq:7}
  \begin{split}
   \widehat{Q_{n+1}}(q_k) &\leftarrow \left(1 + \beta H\hspace{-1mm}\left(\widehat{Q_{n}}(q_{k}); \widehat{Q_{n}}(q_{k-1}), \widehat{Q_{n}}(q_{k+1}) \right) q_k\right) \widehat{Q_{n}}(q_k) \hspace{11mm}\text{ if } \widehat{Q_{n}}(q_k) < x_n \\
   \widehat{Q_{n+1}}(q_k) &\leftarrow \left(1 - \beta H\hspace{-1mm}\left(\widehat{Q_{n}}(q_{k}); \widehat{Q_{n}}(q_{k-1}), \widehat{Q_{n}}(q_{k+1}) \right) (1-q_k)\right) \widehat{Q_{n}}(q_k)\hspace{2mm}\text{ if } \widehat{Q_{n}}(q_k) \geq x_n
  \end{split}
\end{align}
for $k = 2, \ldots, K-1$. For $k=1$ and $k=K$ results into:
\begin{align}
\label{eq:8_2}
  \begin{split}
   \widehat{Q_{n+1}}(q_1) &\leftarrow \left(1 + \beta H\hspace{-1mm}\left(\widehat{Q_{n}}(q_{1}), \widehat{Q_{n}}(q_{2}) \right) q_1\right) \widehat{Q_{n}}(q_1) \hspace{11mm}\text{ if } \widehat{Q_{n}}(q_1) < x_n \\
   \widehat{Q_{n+1}}(q_1) &\leftarrow \left(1 - \beta H\hspace{-1mm}\left(\widehat{Q_{n}}(q_{1}), \widehat{Q_{n}}(q_{2}) \right) (1-q_1)\right) \widehat{Q_{n}}(q_1)\hspace{2mm}\text{ if } \widehat{Q_{n}}(q_1) \geq x_n
  \end{split}
\end{align}
and
\begin{align}
\label{eq:9}
  \begin{split}
   \widehat{Q_{n+1}}(q_K) &\leftarrow \left(1 + \beta H\hspace{-1mm}\left(\widehat{Q_{n}}(q_{K-1}), \widehat{Q_{n}}(q_{K}) \right) q_K\right) \widehat{Q_{n}}(q_K) \hspace{11mm}\text{ if } \widehat{Q_{n}}(q_K) < x_n \\
   \widehat{Q_{n+1}}(q_K) &\leftarrow \left(1 - \beta H\hspace{-1mm}\left(\widehat{Q_{n}}(q_{K-1}), \widehat{Q_{n}}(q_{K}) \right) (1-q_K)\right) \widehat{Q_{n}}(q_K)\hspace{2mm}\text{ if } \widehat{Q_{n}}(q_K) \geq x_n
  \end{split}
\end{align}
The parameter $\beta \in [0,1)$ controls the size of the update when a new sample arrives, and the $H-$ functions ensure that the monotone property will be satisfied in every iteration. In the rest of the paper, we will refer to the method based on \eqref{eq:7} to \eqref{eq:9} as MDUMIQE which is an abbreviation of Multiple DUMIQE.

Now we will present a theorem that catalogues the properties of the estimators $\widehat{Q_{n}}(q_k),\,\, k = 1,2,\ldots, K$ given in \eqref{eq:7} to \eqref{eq:9} for a stationary data stream, i.e. $X_n = X \sim f(x), \,\, n=1,2,\ldots$. We assume that all the estimators $\widehat{Q_{n}}(q_k) > 0,\,\, k = 1,2,\ldots, K$ and the true quantiles $Q(q_k) > 0,\,\, k = 1,2,\ldots, K$. A sufficient condition to obtain $Q(q_k) > 0,\, k = 1,2,\ldots,K$ is that the random variable $X$ only takes positive values. 
\begin{theorem}
\label{thm:positive_quantile}
Let $Q(q_k) = {F_X}^{-1}(q_k),\, k = 1,2,\ldots,K$ be the true quantiles to be estimated and suppose that $Q(q_k)> 0,\, k=1,2,\ldots,K$.
In addition, we suppose that $\widehat{Q_1}(q_k) >0,\, k = 1,2,\ldots,K$. Applying the updating rules \eqref{eq:7} to \eqref{eq:9}, we obtain:
\begin{align*}
\lim_{n \beta \to \infty, \beta \to 0}  \widehat{Q_{n}}(q_k) = Q(q_k),\,\,\,\,\, k=1,2,\ldots,K
\end{align*}
\end{theorem}
\noindent The proof of the theorem can be found in Appendix \ref{app:proof}. Although the quantile estimators $\widehat{Q_{n}}(q_k),\,\, k = 1,2,\ldots, K$ given in \eqref{eq:7} to \eqref{eq:9} are designed to estimate quantiles for a dynamic environment, it is an important quality check of the estimators that they converge to the true quantiles for static data streams as given by Theorem \ref{thm:positive_quantile}.

We end this section with a few remarks.\\[0mm]
\textit{Remark 1}: A potential challenge with multiplicative update schemes, as given by \eqref{eq:1} and by \eqref{eq:7} to \eqref{eq:9}, is that if we start with a quantile estimate above zero, $\widehat{Q_{0}}(q_k) > 0$, the estimates will stay above zero. Similarly, if we start with a quantile estimate below zero, it will stay below zero. A simple solution is to estimate the quantiles of a transformation of the data $h(X_n)$ where $h(\cdot)$ is a monotonically increasing function and $h(x)>0\, \forall x$. A natural alternative is $h(x) = \exp(x)$.\\[1mm]
\textit{Remark 2}: We see that the updating rules \eqref{eq:7} to \eqref{eq:9} only update based on $\widehat{Q_{n}}(q_k) < x_n$ or $\widehat{Q_{n}}(q_k) > x_n$ and not the value of $x_n$. This means that the algorithm is very robust against outliers, which is important for real life data streams.\\[1mm]
\textit{Remark 3}: The strategy of adjusting the value $\lambda$ in order to avoid the monotone violation of quantiles, as described in Section \ref{sec:md}, can also be used for the additive alternative to DUMIQE in \eqref{eq:1} given by
 \begin{align}
\label{eq:1a}
  \begin{split}
   \widehat{Q_{n+1}}(q_k) &\leftarrow \widehat{Q_{n}}(q_k) + \lambda q_k   \hspace{14mm}\text{ if } \widehat{Q_{n}}(q_k) < x_n \\
   \widehat{Q_{n+1}}(q_k) &\leftarrow \widehat{Q_{n}}(q_k) - \lambda (1-q_k)  \,\hspace{5mm}\text{ if } \widehat{Q_{n}}(q_k) \geq x_n
  \end{split}
\end{align}
In our experiments MDUMIQE outperformed this additive alternative and therefore the experimental results in the next section will be related to MDUMIQE.

\section{Experiments}
\label{sec:exper}

In this section, we evaluate the performance of the estimators presented in this paper.
It would be interesting to evaluate the performance of different methods for real life data, but this is challenging to do in a systematic way for dynamical data streams as the ground truth ground truth generally is missing. Before proceeding to systematic experiments based synthetic data, we just recall Figure \ref{fig:6} showing that MDUMIQE can be used to efficiently track quantiles of challenging real life data streams.

We look at two different cases where we assume that the data are outcomes from a normal distribution or from a $\chi^2$ distribution. For the normal distribution case, we assume that the expectation of the distribution varies with time
\begin{align*}
\mu_n = a \sin \left( \frac{2\pi}{T} n \right), \,\,\, n = 1,2,3, \ldots
\end{align*}
which is the sinus function with period $T$. Further, we assume that the standard deviation of the distribution does not vary with time and is equal to one. For the $\chi^2$ distribution case, we assume that the number of degrees of freedom varies with time as follows:
\begin{align*}
  \nu_n = a \sin \left( \frac{2\pi}{T} n \right) + b, \,\,\, n = 1,2,3, \ldots
\end{align*}
where $b > a$ such that $\nu_n > 0$ for all $n$. In the experiments below we used $a = 2$ and $b=6$.

\begin{figure}
  \centering
  \includegraphics[width = \textwidth]{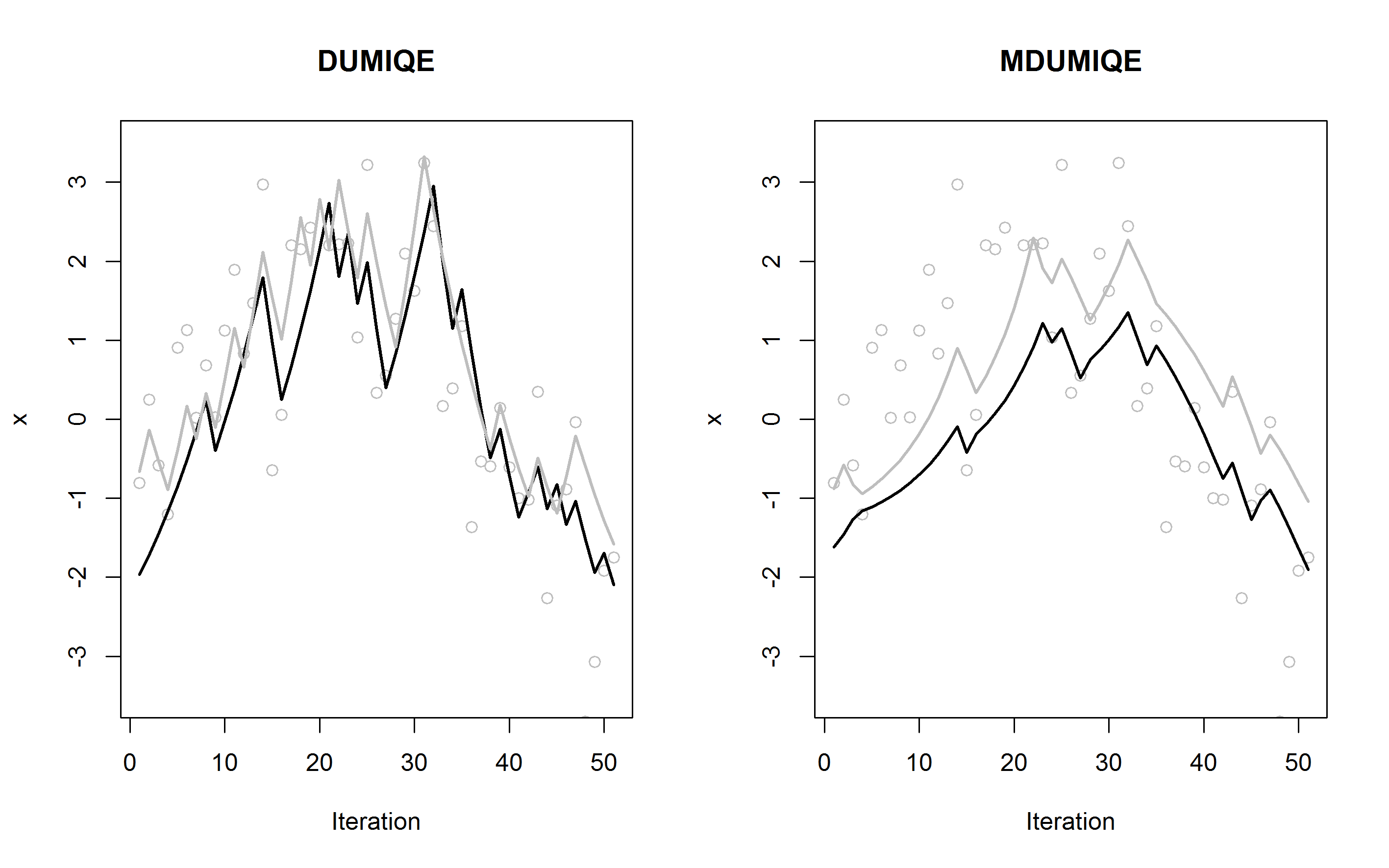}
  \caption{Estimation processes using DUMIQE and MDUMIQE. The gray dots show samples from the data stream distribution while the black and the gray curves show estimates of the 0.4 and the 0.6 quantiles of the data, respectively.}
  \label{fig:7}
\end{figure}

Figure \ref{fig:7} shows a small section of the estimation processes using DUMIQE and MDUMIQUE. The gray dots show the samples from the data stream and is the same in both panels. The data are generated from the normal distribution above with $T=75$. The gray and the black curves show estimates of the 0.4 and the 0.6 quantiles of the data, respectively.
We see that using DUMIQE, the monotone property is violated in several iterations while MDUMIQE satisfies the monotone property and at the same time is able to track the quantiles efficiently.

Now, we turn to conducting a thorough analysis of how well the proposed methods in Section \ref{sec:multq} estimate the quantiles of data streams. We estimated quantiles of both the normally and $\chi^2$ distributed data streams above using two different periods, namely $T=800$ (rapid variation) and $T=8000$ (slow variation), i.e. in total four different data streams. In addition, for each of the four data streams we estimated quantiles that were centred around the median or in the tail of the distribution. We either estimate three or nine quantiles representing cases where the distance between the quaniltes are either large or small, respectively. Obviously, if the quantiles are close to each other, the monotone property will be violated more frequently, making the estimation problem  more difficult. In more detail, for the different cases we estimated the following quantiles:
\begin{itemize}
\item For the normal distribution and the quantiles around the median, we estimated the quantiles related to the following probabilities $q_k = \Phi(-0.8 + 0.2(k-1)), \,\,\, k=1,2,\ldots,9$. For the case with three quantiles, we only used $k=1,5$ and $9$. $\Phi(\cdot)$ refers to the cumulative distribution function of the standard normal distribution. Recall that in dynamically changing data streams, as in these experiments, the value of a quantile related to a specific probability varies with time.
\item For the normal distribution and the quantiles in the tail of the distribution, we use $q_k = \Phi(0.8 + 0.2(k-1)), \,\,\, k=1,2,\ldots,9$. For the case with three quantiles we only used $k=1,5$ and $9$.
\item For the $\chi^2$ distribution and the quantiles around the median, we estimated the quantiles related to the following probabilities $q_k = F(4.2 + 0.3(k-1); \nu = 6), \,\,\, k=1,2,\ldots,9$ where $F(\cdot;\nu)$ refers to the cumulative distribution function of the $\chi^2$ distribution with $\nu$ degrees of freedom. For the case with three quantiles we only used $k=1,5$ and $9$.
\item Finally, for the $\chi^2$ distribution and the quantiles in the tail of the distribution, we estimated the quantiles related to the following probabilities $q_k = F(12 + 0.4(k-1); \nu = 6), \,\,\, k=1,2,\ldots,9$. For the case with three quantiles we only used $k=1,5$ and $9$.
\end{itemize}
The probabilities related to quantiles in the median and in tail of the distribution are  centered around the probabilities $0.5$ and $0.95$, respectively.
When estimating nine quantiles, the choices above resulted in a monotone property violation at about every third iteration using a typical value of $\lambda = 0.05$ in \eqref{eq:1}. Similarly when estimating three quantiles, we got a monotone property violation at about every eleventh iteration.

To measure estimation error, we use the average of the root mean squares error (RMSE) for each quantile given as:
\begin{align}
  \label{eq:27}
  RMSE =  \frac{1}{K} \sum_{k=1}^K  \sqrt{ \frac{1}{N}\sum_{n=1}^N \left(Q_n(q_k) - \widehat{Q_n}(q_k)\right)^2 }
\end{align}
where $N$ is the total number of samples in the data stream. In the experiments, we used $N = 10^7$ which efficiently removes any Monte Carlo errors in the experimental results. In order to get a good overview of the performance of the algorithms, we measure the estimation error for a large set of different values of the tuning parameters $\lambda$ and $\beta$.

\begin{figure}
  \centering
  \includegraphics[width = \textwidth]{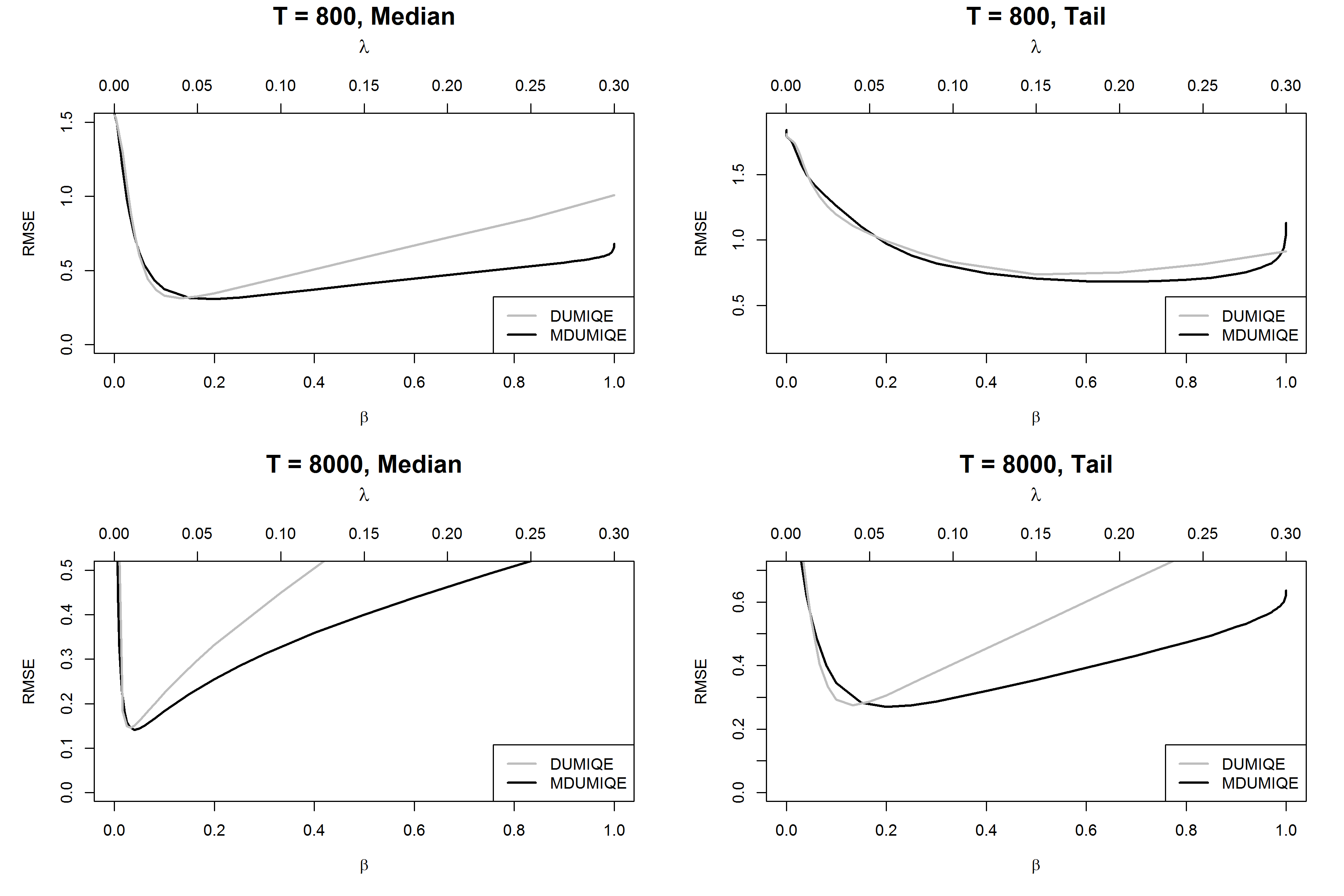}
  \caption{Estimation error for data from the normal distribution when estimating three quantiles.}
  \label{fig:3}
\end{figure}

\begin{figure}
  \centering
  \includegraphics[width = \textwidth]{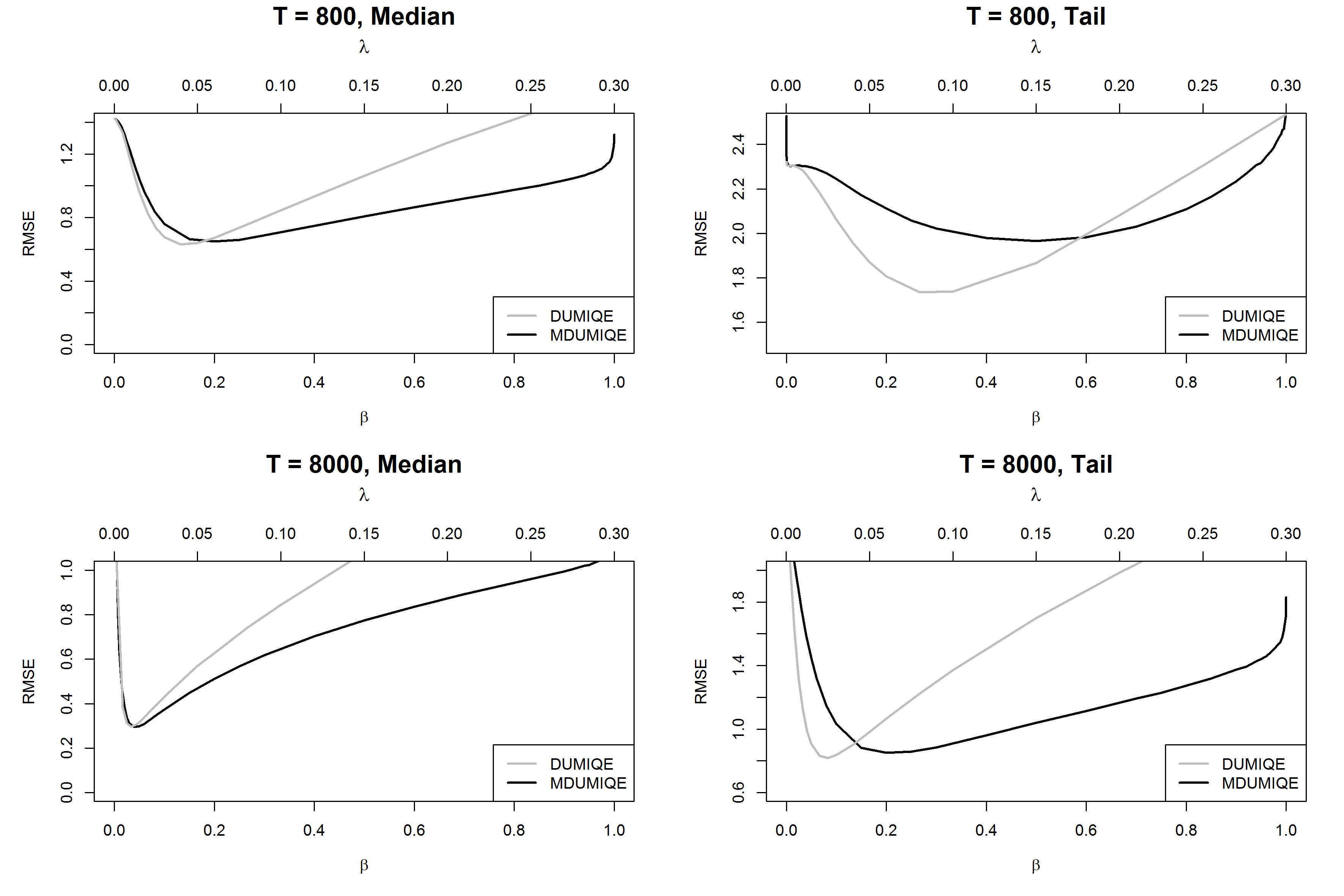}
  \caption{Estimation error for data from the $\chi^2$ distribution when estimating three quantiles.}
  \label{fig:4}
\end{figure}

The results for the normal and $\chi^2$ distribution cases when estimating three quantiles, are shown in Figures \ref{fig:3} and \ref{fig:4}, respectively.

We proceed to discussing the normal distribution cases. For all the four cases, we observe that MDUMIQE outperforms DUMIQE for the optimal values of $\beta$ and $\lambda$ are chosen (results in the smallest RMSE). We also see that estimation performance of the MDUMIQE is less sensitive to the choice of $\beta$ than DUMIQE on the choice of $\lambda$. This is a crucial remark since for real life applications we do not know the optimal values of $\beta$ and $\lambda$ that yield the best results. Hence, not only are we able to satisfy the monotone property of quantiles, we also improve estimation precision compared to DUMIQE. For the $\chi^2$ distribution cases we see that MDUMIQE and DUMIQE performs about equally well except that DUMIQE performs better when $T = 800$ and when estimating quantiles in the tail of the distribution. The reason will be explained below.

\begin{figure}
  \centering
  \includegraphics[width = \textwidth]{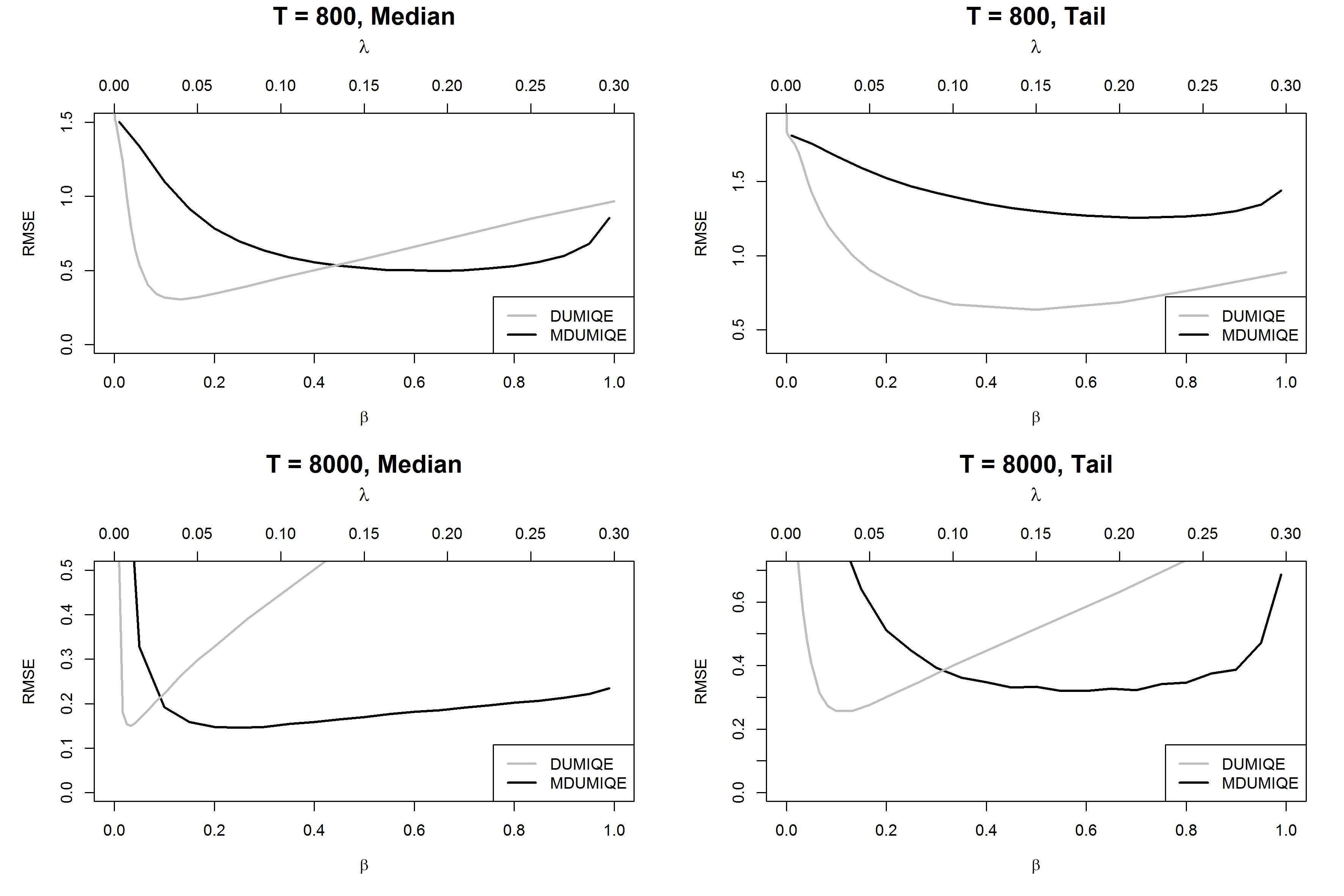}
  \caption{Estimation error for data from the normal distribution when estimating nine quantiles.}
  \label{fig:5}
\end{figure}

\begin{figure}
  \centering
  \includegraphics[width = \textwidth]{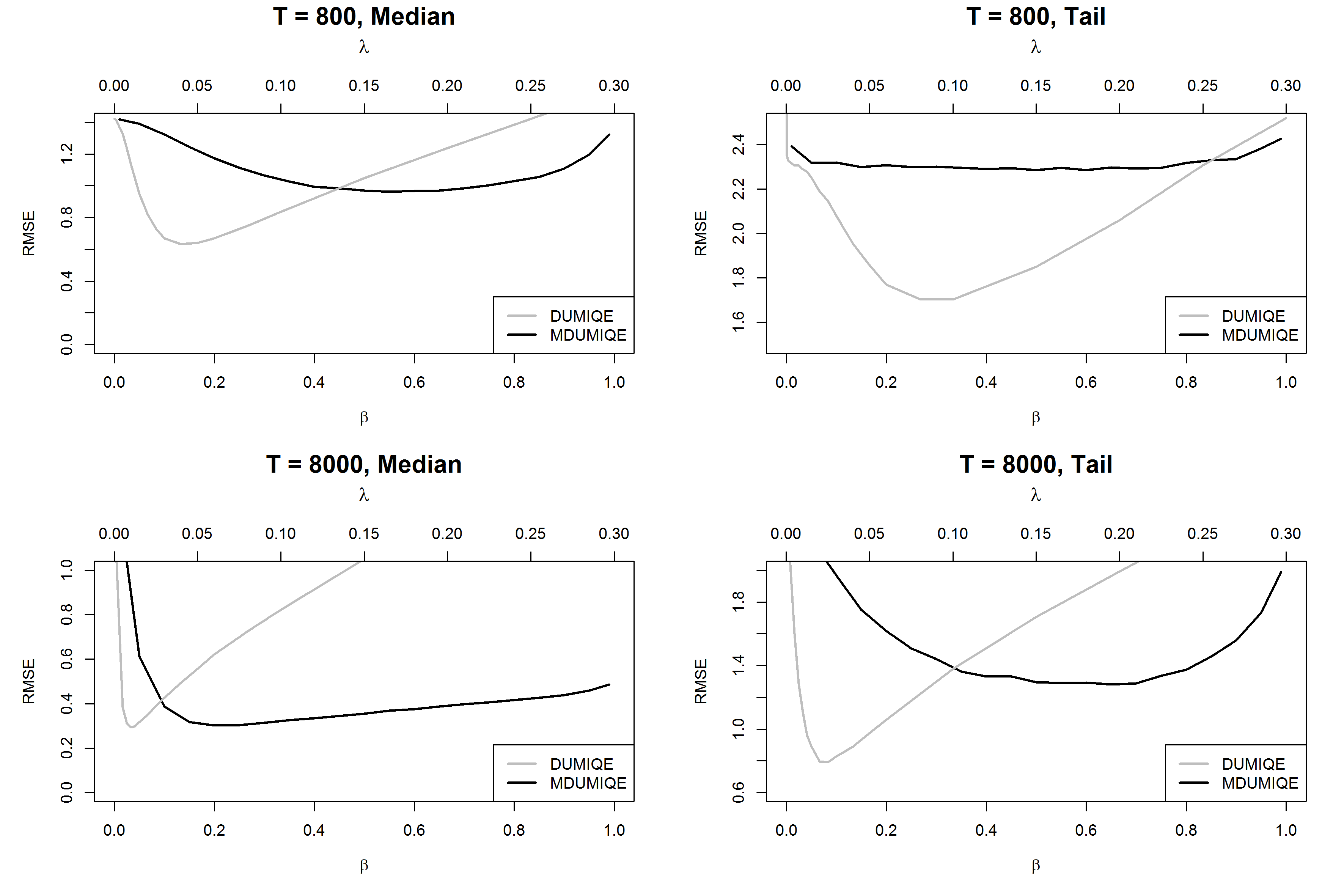}
  \caption{Estimation error for data from the $\chi^2$ distribution when estimating nine quantiles.}
  \label{fig:2}
\end{figure}

The results for the normal and $\chi^2$ distribution cases when estimating nine quantiles are shown in Figures \ref{fig:5} and \ref{fig:2}, respectively.
We start by discussing the normal distribution cases. We see that for $T = 800$, the DUMIQE performs better than the MDUMIQE, especially when estimating in the tail of the distribution. On the other hand, for $T=8000$ and estimating the median, the MDUMIQE outperforms DUMIQE. The explanation for why DUMIQE performs better when $T=800$ is that for such a rapidly changing data stream, large updates of the quantile estimates must be used to track the true quantiles. MDUMIQE is required to satisfy the monotone property which sets a limitation on how far MDUMIQE can update the estimates in each iteration. More specifically, for MDUMIQE, the estimate $\widehat{Q_{n}}(q_{k})$ must always be between $\widehat{Q_{n}}(q_{k-1})$ and $\widehat{Q_{n}}(q_{k+1})$, recall \eqref{eq:6}. Whenever the difference between $\widehat{Q_{n}}(q_{k-1})$ and $\widehat{Q_{n}}(q_{k+1})$ is small,  we can only do small updates of $\widehat{Q_{n}}(q_{k})$.
For the $\chi^2$ distribution, the two methods perform about equally well for $T=8000$ and estimating the median and for the other cases DUMIQE performs better MDUMIQE. An appealing property of the MDUMIQE approach (in addition to satisfying the monotone property) is that the estimation performance is less sensitive to the choice of $\beta$ than DUMIQE is on the choice of $\lambda$. Using $\beta = 0.5$ we get satisfactory results in all the cases. Such a ``universal'' $\lambda$ do not exist for DUMIQE.

For comparison purposes, we also tested the multiple quantile estimation method in \cite{cao2009incremental} for the estimation tasks described above. This is the only viable method we have found in the literature for estimating multiple quantiles in dynamically changing data streams. The method has two tuning parameters, a weight parameter similar to $\lambda$ and $\beta$ for the methods in this paper, and a parameter that controls the width of intervals to estimate the distribution of the data stream around a quantile. To achieve the best possible results we ran the method for a large set of values for the two parameters. The best estimation results (smallest RMSE) are shown in Tables \ref{tab:1} and \ref{tab:2} for the cases with three and nine quantiles, respectively.
\begin{table}
  \centering
  \begin{tabular}{lcccc}
    & $T = 800$, Median & $T = 800$, Tail & $T = 8000$, Median & $T = 8000$, Tail  \\ \hline
Normal distribution & 0.835 & 1.00 & 0.223 & 0.570  \\[3mm]
    & $T = 800$, Median & $T = 800$, Tail & $T = 8000$, Median & $T = 8000$, Tail  \\ \hline
$\chi^2$ distribution & 1.512 & 3.93 & 1.00 &  3.75\\
  \end{tabular}
  \caption{Estimation error (RMSE using \eqref{eq:27}) using the method in Cao et al. (2009) \cite{cao2009incremental} when estimating three quantiles.}
  \label{tab:1}
\end{table}
\begin{table}
  \centering
  \begin{tabular}{lcccc}
    & $T = 800$, Median & $T = 800$, Tail & $T = 8000$, Median & $T = 8000$, Tail  \\ \hline
Normal distribution & 0.312 & 0.630 & 0.259 & 0.370  \\[3mm]
    & $T = 800$, Median & $T = 800$, Tail & $T = 8000$, Median & $T = 8000$, Tail  \\ \hline
$\chi^2$ distribution & 0.79 & 2.40 & 0.445 & 1.611 \\
  \end{tabular}
  \caption{Estimation error (RMSE using \eqref{eq:27}) using the method in Cao et al. (2009) \cite{cao2009incremental} when estimating nine quantiles.}
  \label{tab:2}
\end{table}
Comparing these results with the results in Figures \ref{fig:3} to \ref{fig:2}, we see that MDUMIQE clearly outperforms Cao et al. (2009) \cite{cao2009incremental}.

\section{Closing remarks}

In this paper, we present a novel algorithm for estimating multiple quantiles in a dynamically changing data stream.
The algorithm is an extension of the efficient DUMIQE from \cite{yazidi16}, developed to avoid monotone violations of the quantile estimates.

The experimental results in Section \ref{sec:exper} show that the suggested algorithm performs very well. For most of the experiments the method performs equally well or better than DUMIQE that do not satisfy the monotone property of quantiles. The suggested algorithm is well suited to track multiple quantiles for real life data streams as it adapts efficiently to dynamic changes in the data streams and is robust to outliers.

Another advantage of the suggested algorithm is that the estimation performance is less sensitive to the choice of the tuning parameter compared to DUMIQE. Choosing a $\beta = 0.5$ performed fairly well in all the experiments. This is a crucial property for real life data streams since the distribution of data streams may vary slowly in some time periods and more rapidly in others, see Figure \ref{fig:6} for an example. Using the DUMIQE one must choose a tuning parameter that performs well either where the data stream varies slowly or rapidly. Since the performance of the algorithm suggested in this paper is less sensitive to the choice of the tuning parameter, it will perform well both when the data stream varies slowly and rapidly. In Figure \ref{fig:6} we see that the algorithm performs well both when the data stream varies slowly and rapidly. In addition, we saw that the algorithm outperformed the state of the art method of Cao et al. \cite{cao2009incremental}.

The suggested algorithm experiences some reduction in performance for rapidly changing data streams. For such data streams, large updates of the quantile estimates are necessary to track the true quantiles efficiently. The requirement of satisfying the monotone property of quantiles sets a limit on how large updates that are possible, and thus reduces the performance of the algorithm. An interesting challenge for future research, is to develop an incremental quantile estimation algorithm that performs better in rapidly changing dynamically changing data streams when many quantiles need to be estimated.

\bibliographystyle{plain}
\bibliography{bibl}

\clearpage

\appendix
\section{Proof of Theorem \ref{thm:positive_quantile}}
\label{app:proof}

We will first present a theorem due to Norman \cite{Norman1972} that will be used to prove Theorem \ref{thm:positive_quantile}.
Norman \cite{Norman1972} studied distance "diminishing models". The convergence of $\widehat{Q_n}(q_k)$ to $Q(q_k)$ is a consequence of this theorem.
\begin{theorem}
\label{thm:Norman}
Let $x(t)$  be a stationary Markov process dependent on a constant parameter $\theta \in [0,1]$. Each $x(t) \in I$, where $I$ is  a  subset  of  the  real  line.  Let $\delta x(t)=x(t+1)-x(t)$. The following are assumed to hold:
\begin{enumerate}
\item I is compact
\item $E [\delta x(t) | x(t)=y]= \theta w(y)+ O(\theta^2)$
\item $Var [\delta x(t) | x(t)=y]= \theta ^2 s(y)+ O(\theta^2)$
\item $E [\delta x(t)^3 | x(t)=y]=  O(\theta ^3)$
where $sup_{y \in I} \frac{O(\theta^k)}{\theta^k}< \infty$ for $k=2,3$ and $sup_{y \in I} \frac{o(\theta^2)}{\theta^2} \rightarrow 0$ as $\theta \rightarrow 0$
\item $w(y)$ has a Lipschitz derivative in $I$
\item $s(y)$ is Lipschitz $I$.
\end{enumerate}
If Assumptions 1 to 6 above hold, $w(y)$ has a unique root $y^*$ in $I$ and
$\frac{d w}{d y}  \bigg|_{y=y^*} \le 0$ then
\begin{enumerate}
\item $var [\delta x(t) | x(0)=x]=O(\theta)$ uniformly for all $x \in I$ and $t \ge 0$.
For any $x \in I$,  the differential equation $\frac{d y(\tau)}{d \tau}=w(y(t))$ has a unique solution  $y(\tau)=y(\tau,x)$ with $y(0)=x$  and	$E [\delta x(t) | x(0)=x]=y( t \theta)+O(\theta)$ uniformly for all $x \in I$ and  $t \ge 0$.
\item $\frac{x(t)-y(t \theta)}{\sqrt \theta}$ has a normal distribution with zero mean and finite variance as $\theta \rightarrow 0$ and $t \theta \rightarrow \infty$.
\end{enumerate}
\end{theorem}
Having presented Theorem \ref{thm:Norman}, we are now ready to prove Theorem \ref{thm:positive_quantile} by resorting to Theorem \ref{thm:Norman}. This is the main result of this paper .
We will prove the convergence of $\widehat{Q_n}(q_k)$ for $k=2,3,\ldots,K-1$ below. The proof for $\widehat{Q_n}(q_1)$ and $\widehat{Q_n}(q_K)$ can be done in the same manner and are not shown in this paper for the sake of brevity.
\begin{proof}
We now start by showing that the Markov process based on the updating rules \eqref{eq:7} to \eqref{eq:9} satisfies the assumptions 1 to 6 in Theorem \ref{thm:Norman}.
\begin{align}
\notag
  &E\left(\delta \widehat{Q_{n}}(q_k)\,\left|\,\widehat{Q_n}(q_k)\right.\right) = \\
\notag
  & = E\left(\delta \widehat{Q_{n}}(q_k)\,\left|\,\widehat{Q_n}(q_k) \geq X\right. \right) P\left(\widehat{Q_n}(q_k) \geq X \right) + E\left(\delta \widehat{Q_{n}}(q_k)\,|\,\widehat{Q_n}(q_k) < X\right) P\left(\widehat{Q_n}(q_k) < X\right) = \\
\notag
  & = \beta H\hspace{-1mm}\left(\widehat{Q_{n}}(q_{k}); \widehat{Q_{n}}(q_{k-1}), \widehat{Q_{n}}(q_{k+1}) \right) q_k \widehat{Q_{n}}(q_{k})\left(1 - F_X\left(\widehat{Q_{n}}(q_{k})\right)\right) \\
\notag
  & - \beta H\hspace{-1mm}\left(\widehat{Q_{n}}(q_{k}); \widehat{Q_{n}}(q_{k-1}), \widehat{Q_{n}}(q_{k+1}) \right) (1 - q_k) \widehat{Q_{n}}(q_{k}) F_X\left(\widehat{Q_{n}}(q_{k})\right) \\
\label{eq:10}
  &= \beta H\hspace{-1mm}\left(\widehat{Q_{n}}(q_{k}); \widehat{Q_{n}}(q_{k-1}), \widehat{Q_{n}}(q_{k+1}) \right) \widehat{Q_{n}}(q_{k}) \left(q_k - F_X\left(\widehat{Q_{n}}(q_{k})\right)\right)
\end{align}
We now let $\theta = \beta$, $y = \widehat{Q_{n}}(q_{k})$ and $w\hspace{-0.5mm}\left(\widehat{Q_{n}}(q_{k})\right)$ be equal to "everything" in \eqref{eq:10} except $\beta$. It is easy to see that assumption 2 in Theorem \ref{thm:Norman} is satisfied. Next, we turn to assumption 5 which requires that $w\left(\widehat{Q_{n}}(q_{k})\right)$ has a Lipschitz derivative with respect to $\widehat{Q_{n}}(q_{k})$. Unfortunately it is not obvious that this is satisfied since $H$ has a discontinuous derivative with respect to $\widehat{Q_{n}}(q_{k})$ due to the min-function in \eqref{eq:6}. To show that both assumptions 2 and 5 are satisfied, we need to perform a subtle modification of \eqref{eq:10} as follows. A typical example of $H$ as a function of $\widehat{Q_{n}}(q_{k})$ is shown as the black curve in Figure \ref{fig:1}.
  \begin{figure}
    \centering
    \includegraphics[width = 0.8\textwidth]{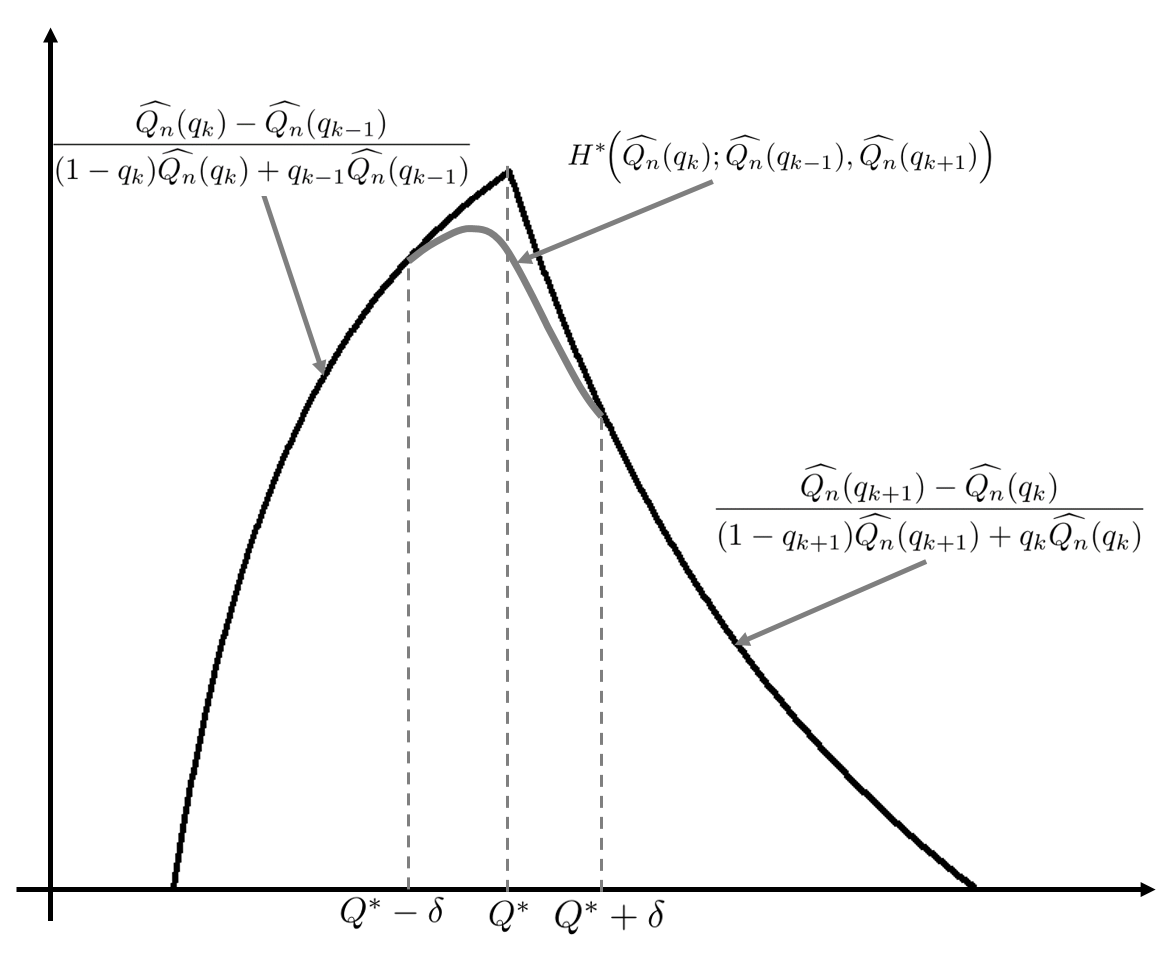}
    \caption{The black and the gray curves show the functions $H$ and $H^{\ast}$, respectively.}
    \label{fig:1}
  \end{figure}
We define a function $H^{\ast}$ that is equal to $H$ except for an the interval $[Q^{\ast} - \delta, Q^{\ast} + \delta]$ (see Figure \ref{fig:1}). In the interval $[Q^{\ast} - \delta, Q^{\ast} + \delta]$, $H^{\ast}$ is a function that is smaller then $H$ (to satisfy the monotone property) and has a Lipschitz derivative. This requires that $H^{\ast}$ satisfy the following
\begin{align*}
  &H^{\ast}\hspace{-1mm}\left(Q^{\ast} - \delta; \widehat{Q_{n}}(q_{k-1}), \widehat{Q_{n}}(q_{k+1}) \right) = H\hspace{-1mm}\left(Q^{\ast} - \delta; \widehat{Q_{n}}(q_{k-1}), \widehat{Q_{n}}(q_{k+1}) \right) \\
  &\frac{d H^{\ast}}{d \widehat{Q_{n}}(q_{k})} \left(Q^{\ast} - \delta; \widehat{Q_{n}}(q_{k-1}), \widehat{Q_{n}}(q_{k+1}) \right) = \frac{d H^{\ast}}{d \widehat{Q_{n}}(q_{k})} \left(Q^{\ast} - \delta; \widehat{Q_{n}}(q_{k-1}), \widehat{Q_{n}}(q_{k+1}) \right) \\
  &H^{\ast}\hspace{-1mm}\left(Q^{\ast} + \delta; \widehat{Q_{n}}(q_{k-1}), \widehat{Q_{n}}(q_{k+1}) \right) = H\hspace{-1mm}\left(Q^{\ast} + \delta; \widehat{Q_{n}}(q_{k-1}), \widehat{Q_{n}}(q_{k+1}) \right) \\
  &\frac{d H^{\ast}}{d \widehat{Q_{n}}(q_{k})} \left(Q^{\ast} + \delta; \widehat{Q_{n}}(q_{k-1}), \widehat{Q_{n}}(q_{k+1}) \right) = \frac{d H^{\ast}}{d \widehat{Q_{n}}(q_{k})} \left(Q^{\ast} + \delta; \widehat{Q_{n}}(q_{k-1}), \widehat{Q_{n}}(q_{k+1}) \right)
\end{align*}
i.e. that the function value and the derivative must be equal for $H$ and $H^{\ast}$ in $Q^{\ast} - \delta$ and $Q^{\ast} + \delta$. It is straight forward to satisfy these criteria, e.g. by fitting a polynomial. $H^{\ast}$ is illustrated as the gray curve in Figure \ref{fig:1} (and is equal to $H$ outside the interval). By reducing the value of $\delta$, $H^{\ast}$ will be more and more similar to $H$. In other words, there exists always a $\delta$ such that
\begin{align*}
\left| H\hspace{-1mm}\left(\widehat{Q_{n}}(q_{k}); \widehat{Q_{n}}(q_{k-1}), \widehat{Q_{n}}(q_{k+1}) \right) - H^{\ast}\hspace{-1mm}\left(\widehat{Q_{n}}(q_{k}); \widehat{Q_{n}}(q_{k-1}), \widehat{Q_{n}}(q_{k+1}) \right)\right| < \beta \,\,\,\, \forall\,\widehat{Q_{n}}(q_{k}) \in \left[\widehat{Q_{n}}(q_{k-1}), \widehat{Q_{n}}(q_{k+1}) \right]
\end{align*}
which means that we can write:
\begin{align}
  \label{eq:11}
  H\hspace{-1mm}\left(\widehat{Q_{n}}(q_{k}); \widehat{Q_{n}}(q_{k-1}), \widehat{Q_{n}}(q_{k+1}) \right) = H^{\ast}\hspace{-1mm}\left(\widehat{Q_{n}}(q_{k}); \widehat{Q_{n}}(q_{k-1}), \widehat{Q_{n}}(q_{k+1}) \right) + O(\beta)
\end{align}
Substituting \eqref{eq:11} into \eqref{eq:10} we obtain:
\begin{align*}
  &\beta H^{\ast} \hspace{-1mm}\left(\widehat{Q_{n}}(q_{k}); \widehat{Q_{n}}(q_{k-1}), \widehat{Q_{n}}(q_{k+1}) \right) \widehat{Q_{n}}(q_{k}) \left(q_k - F_X\left(\widehat{Q_{n}}(q_{k})\right)\right) + \beta \widehat{Q_{n}}(q_{k})  \left(q_k - F_X\left(\widehat{Q_{n}}(q_{k})\right)\right) O(\beta) \\
  & = \beta H^{\ast} \hspace{-1mm}\left(\widehat{Q_{n}}(q_{k}); \widehat{Q_{n}}(q_{k-1}), \widehat{Q_{n}}(q_{k+1}) \right) \widehat{Q_{n}}(q_{k}) \left(q_k - F_X\left(\widehat{Q_{n}}(q_{k})\right)\right) + O(\beta^2)
\end{align*}
Since $H^{\ast}$ has a Lipschitz derivative, we see that with:
\begin{align}
  \label{eq:14}
w\left(\widehat{Q_{n}}(q_{k})\right) = H^{\ast}\hspace{-1mm}\left(\widehat{Q_{n}}(q_{k}); \widehat{Q_{n}}(q_{k-1}), \widehat{Q_{n}}(q_{k+1}) \right) \widehat{Q_{n}}(q_{k}) \left(q_k - F_X\left(\widehat{Q_{n}}(q_{k})\right)\right)
\end{align}
both assumptions 2 and 5 are satisfied.

Next we turn to assumption 3.
\begin{align*}
  &E\left(\delta \widehat{Q_{n}}(q_k)^2\,\left|\,\widehat{Q_n}(q_k)\right.\right) = \\
  & = E\left(\delta \widehat{Q_{n}}(q_k)^2\,\left|\,\widehat{Q_n}(q_k) \geq X \right.\right) P\left(\widehat{Q_n}(q_k) \geq X \right) + E\left(\delta \widehat{Q_{n}}(q_k)^2\,\left|\,\widehat{Q_n}(q_k) < X\right.\right) P\left(\widehat{Q_n}(q_k) < X\right) = \\
  & = \beta^2 \left( H\hspace{-1mm}\left(\widehat{Q_{n}}(q_{k}); \widehat{Q_{n}}(q_{k-1}), \widehat{Q_{n}}(q_{k+1}) \right) q_k \widehat{Q_{n}}(q_{k}) \right)^2 \left(1 - F_X\left(\widehat{Q_{n}}(q_{k})\right)\right) \\
  & - \beta^2 \left(H\hspace{-1mm}\left(\widehat{Q_{n}}(q_{k}); \widehat{Q_{n}}(q_{k-1}), \widehat{Q_{n}}(q_{k+1}) \right) (1 - q_k) \widehat{Q_{n}}(q_{k}) \right)^2 F_X\left(\widehat{Q_{n}}(q_{k})\right)
\end{align*}
\begin{align}
  \label{eq:13}
  \begin{split}
  & Var\left(\delta \widehat{Q_{n}}(q_k)\,\left|\,\widehat{Q_n}(q_k)\right.\right) = E\left(\delta \widehat{Q_{n}}(q_k)^2\,\left|\,\widehat{Q_n}(q_k)\right.\right) - E\left(\delta \widehat{Q_{n}}(q_k)\,\left|\,\widehat{Q_n}(q_k)\right.\right)^2 = \\
  & = \beta^2 \left( H\hspace{-1mm}\left(\widehat{Q_{n}}(q_{k}); \widehat{Q_{n}}(q_{k-1}), \widehat{Q_{n}}(q_{k+1}) \right) q_k \widehat{Q_{n}}(q_{k}) \right)^2 \left(1 - F_X\left(\widehat{Q_{n}}(q_{k})\right)\right) \\
  & - \beta^2 \left(H\hspace{-1mm}\left(\widehat{Q_{n}}(q_{k}); \widehat{Q_{n}}(q_{k-1}), \widehat{Q_{n}}(q_{k+1}) \right) (1 - q_k) \widehat{Q_{n}}(q_{k}) \right)^2 F_X\left(\widehat{Q_{n}}(q_{k})\right) \\
  & + \beta^2 \left( H\hspace{-1mm}\left(\widehat{Q_{n}}(q_{k}); \widehat{Q_{n}}(q_{k-1}), \widehat{Q_{n}}(q_{k+1}) \right) \widehat{Q_{n}}(q_{k}) \left(q_k - F_X\left(\widehat{Q_{n}}(q_{k})\right)\right)\right)^2
  \end{split}
\end{align}
We see that assumption 3 is satisfied with $s\hspace{-0.7mm}\left(\widehat{Q_{n}}(q_{k})\right)$ equal to everything in \eqref{eq:13} except $\beta^2$. Since $H$ is Lipschitz (has a bounded derivative), $s\hspace{-0.7mm}\left(\widehat{Q_{n}}(q_{k})\right)$ is Lipschitz and assumption 6 is also satisfied. Assumption 4 can now be proved in the same manner.

We will use the results of Norman to prove the convergence. It is easy to see that $w\left(\widehat{Q_{n}}(q_{k})\right)$ in \eqref{eq:14} admits two roots $\widehat{Q_n}(q_k) = {F_X}^{-1}(q_k) = Q(q_k)$ and $\widehat{Q_n}(q_k) = 0$.
By introducing an arbitrarily small lower bound $Q_{\text{min}}>0$ on estimate $\widehat{Q_{n}}(q_{k})$, we can avoid the $\widehat{Q_n}(q_k)=0$.
This is easily implemented by modifying the update rules and adding $Q_{\text{min}}$ to the right term of equations (\ref{eq:7}) to (\ref{eq:9}). Therefore the unique root becomes $\widehat{Q_n}(q_k)={F_X}^{-1}(q_k) = Q(q_k)$.

We now differentiate to get:
\begin{align*}
\frac{d\, w\left(\widehat{Q_{n}}(q_{k})\right)}{d\, \widehat{Q_{n}}(q_{k}) } &=  \left( \frac{d\, H^{\ast}}{d\, \widehat{Q_{n}}(q_{k})} + H^{\ast} \right)(q_k - F_X\left( \widehat{Q_{n}}(q_{k}) \right) - H^{\ast} \widehat{Q_{n}}(q_{k}) f\left(\widehat{Q_{n}}(q_{k})\right)
\end{align*}
We substitute the unique root $Q(q_k)$ for $\widehat{Q_{n}}(q_{k})$ and get
\begin{align*}
\frac{ d\, w\left(\widehat{Q_{n}}(q_{k})\right) }{d\, \widehat{Q_{n}}(q_{k})} \bigg|_{\widehat{Q_{n}}(q_{k})=Q(q_k)} &=  \left( \frac{d\, H^{\ast}}{d\, \widehat{Q_{n}}(q_{k})} + H^{\ast} \right)(q_k - F_X\left( Q(q_{k}) \right) - H^{\ast} Q(q_{k}) f\left(Q(q_{k})\right)  \\
&=  0-H^{\ast} Q(q_k) f_X(Q(q_k))<0.
\end{align*}
This gives
\begin{align*}
  \lim_{n \beta  \to \infty, \beta \to 0}  E\left(\widehat{Q_n}(q_k)\right)=Q(q_k)+O(\beta)
\end{align*}
and
\begin{align*}
Var\left(\widehat{Q_n}(q_k)\right)=O(\beta)
\end{align*}
Consequently
\begin{align*}
  \lim_{n \beta  \to \infty, \beta \to 0}  \widehat{Q_n}(q_k)=Q(q_k)
\end{align*}
\end{proof}

\end{document}